\documentclass[11pt]{article}
\usepackage{epsfig} 
\newcommand{\half}{\mbox{\small{$\frac{1}{2}$}}} 
\newcommand{\MSbar}{\overline{\mbox{MS}}} 

\begin{document}
\title{Chiral exponents in $O(N)$ $\times$ $O(m)$ spin models at $O(1/N^2)$} 
\author{J.A. Gracey, \\ Theoretical Physics Division, \\ Department 
of Mathematical Sciences, \\ University of Liverpool, \\ Peach Street, \\ 
Liverpool, \\ L69 7ZF, \\ United Kingdom.} 
\date{} 
\maketitle 
\vspace{5cm} 
\noindent 
{\bf Abstract.} The critical exponents corresponding to chirality are computed
at $O(1/N^2)$ in $d$-dimensions at the stable chiral fixed point of a scalar
field theory with an $O(N)$~$\times$~$O(m)$ symmetry. Pad\'{e}-Borel estimates 
for the exponents are given in three dimensions for the Landau-Ginzburg-Wilson
model at $m$ $=$ $2$. 

\vspace{-16cm} 
\hspace{13.5cm} 
{\bf LTH 551} 

\newpage 
Scalar quantum field theories have been widely used to determine information at
criticality for a large range of critical phenomena. The primary tool of 
computation is the renormalization group equation where the field theory is 
renormalized in perturbation theory to a certain loop order. The resulting 
renormalization group functions are then evaluated at the appropriate fixed 
point, such as that of Wilson and Fisher, \cite{1}, and a numerical estimate 
provided for the corresponding critical exponents, \cite{2,3}. A recent, 
comprehensive and detailed review of the use of the renormalization group in 
this area is given in \cite{4}. For the much studied Heisenberg system, the 
$O(N)$ $\phi^4$ theory has been renormalized to five loops in conventional 
dimensionally regularized perturbation theory, \cite{5} and references therein,
and to six loops in \cite{6} where the spacetime dimension is fixed to be 
three. The resulting critical exponents after appropriate resummation are in 
very reasonable agreement with experiment and other methods such as Monte Carlo 
simulations and the high temperature series expansion, demonstrating the 
success of the approach. More recently, there has been renewed interest in 
understanding the phase transitions in spin models which are similar to the 
usual Heisenberg spin systems. These are spin systems where there is an 
additional chiral symmetry which gives rise to fixed points in the system over 
and above the usual Gaussian free field and Heisenberg fixed points. In other 
words there is a chiral transition in systems with noncollinear order. Early 
work on this system was carried out by Kawamura in \cite{7,8,9,10}. From the 
field theoretic point of view the underlying quantum field theory of such a 
spin system is a $\phi^4$ model with an $O(N)$~$\times$~$O(m)$ symmetry where 
to relate to the chiral property one sets $m$~$=$~$2$ thereby reducing it to a 
Landau-Ginzburg-Wilson theory. However, the more general models have been 
studied for arbitrary $m$ to try and unravel the true fixed point structure of 
the field theory. This is important as there appears to be conflicting evidence
from theoretical calculations, either using Monte Carlo techniques or 
renormalization group methods, as to the number of stable fixed points and 
their existence and as to whether they are first or second order. Prior to 
\cite{11} the underlying field theory had only been renormalized to two loops 
in the $\MSbar$ scheme. In \cite{11} this was extended to three loops and the 
fixed dimension calculation of critical exponents was performed to six loops in
\cite{12}. Moreover, in \cite{13} the chiral exponents associated with the 
stable chiral fixed point of the $O(N)$~$\times$~$O(m)$ model have been 
computed to six loops in three dimensions. They have been shown to be in good 
agreement with other results, \cite{10,14,15}. Briefly the method of 
calculation is to compute the anomalous dimension of a chiral bilinear operator
and evaluate it at the stable chiral fixed point when $m$~$=$~$2$. 

Aside from explicit perturbation theory there are other field theory methods 
with which one can estimate critical exponents. One such method is the large 
$N$ expansion where the exponents are determined order by order in powers of 
$1/N$. The critical point large $N$ technique developed in \cite{16,17} has 
proved to be an excellent tool to compute critical exponents. For instance, in 
the $O(N)$ $\phi^4$ theory the wave function exponent or anomalous dimension, 
$\eta$, is known at $O(1/N^3)$ in arbitrary spacetime dimension $d$, \cite{18}. 
Through the critical renormalization group equation the coefficients of the 
$\epsilon$-expansion of such exponents are in one-to-one agreement with the 
coefficients of the corresponding perturbative renormalization group functions 
evaluated at the same fixed point, where $d$ $=$ $4$ $-$ $2\epsilon$. Given the
recent interest in the $O(N)$~$\times$~$O(m)$ field theory it is the purpose of
this article to calculate the same chiral exponents of \cite{13} in the large
$N$ expansion at $O(1/N^2)$ in $d$-dimensions using the approach of 
\cite{16,17,19}. Indeed in \cite{11} the exponents $\eta$ and $\nu$ have 
already been computed at $O(1/N^2)$ in $d$-dimensions at each fixed point. Our 
calculation involves determining the critical exponent corresponding to the
anomalous dimension of the chiral operator and parallels a similar calculation 
in the $O(N)$ $\phi^4$ theory itself where the critical exponent of the 
bilinear operator corresponding to crossover behaviour was determined at 
$O(1/N^2)$, \cite{20}. 

The (massless) Lagrangian underlying the chiral phase transition in frustrated 
spin systems with noncollinear order is based on that for the
Landau-Ginzburg-Wilson $O(N)$~$\times$~$O(2)$ model. Therefore, we take 
\begin{equation} 
L ~=~ \frac{1}{2} \partial^\mu \phi^{a\alpha} \partial_\mu \phi^{a\alpha} ~+~ 
\frac{\bar{g}}{4!} \left( \phi^{a\alpha} \phi^{a\alpha} \right)^2 ~+~ 
\frac{g_T}{4!} \left[ \left( \phi^{a\alpha} \phi^{a\beta} \right)^2 ~-~ 
\left( \phi^{a\alpha} \phi^{a\alpha} \right)^2 \right] 
\label{lag1} 
\end{equation}  
where $1$ $\leq$ $a$ $\leq$ $N$, $1$ $\leq$ $\alpha$ $\leq$ $m$ and $\bar{g}$ 
and $g_T$ are independent coupling constants. Alternatively, it can be 
reformulated in terms of the auxiliary fields $\sigma$ and $T^{\alpha\beta}$ 
by, \cite{11},  
\begin{equation}
L ~=~ \frac{1}{2} \partial^\mu \phi^{a\alpha} \partial_\mu \phi^{a\alpha} ~+~ 
\frac{1}{2} \sigma \phi^{a\alpha} \phi^{a\alpha} ~+~ 
\frac{1}{2} T^{\alpha\beta} \phi^{a\alpha} \phi^{a\beta} ~-~ 
\frac{3\sigma^2}{2g} ~-~ \frac{3T^{\alpha\beta} T^{\alpha\beta}}{2g_T} 
\label{lag2} 
\end{equation} 
where in our notation $\bar{g}$ $=$ $g$ $+$ $(m-1)g_T/m$ and $T^{\alpha\beta}$ 
is an $O(m)$ symmetric and traceless auxiliary field. Following the treatment 
of \cite{8,11} there are in fact four fixed points in (\ref{lag2}) which are 
the usual Gaussian and Heisenberg ones together with the stable chiral and 
unstable antichiral fixed points. For the purposes of this article we are 
interested in the stable chiral fixed point. In the large $N$ critical point 
approach of \cite{16,17} the propagators of the fields of (\ref{lag2}) take the
following asymptotic scaling forms in the critical region in momentum space  
\begin{equation}
\phi(k) ~ \sim ~ \frac{A}{(k^2)^{\mu-\alpha}} ~~~,~~~ 
\sigma(k) ~ \sim ~ \frac{B}{(k^2)^{\mu-\beta}} ~~~,~~~  
T^{\alpha\beta,\sigma\rho}(k) ~=~ X^{\alpha\beta,\sigma\rho} T(k) 
\label{props} 
\end{equation} 
where 
\begin{equation} 
T(k) ~ \sim ~ \frac{C}{(k^2)^{\mu-\gamma}} 
\label{propT}
\end{equation} 
and
\begin{equation} 
X^{\alpha\beta,\sigma\rho} ~=~ \frac{1}{2} \left( \delta^{\alpha\sigma} 
\delta^{\beta\rho} ~+~ \delta^{\alpha\rho} \delta^{\beta\sigma} ~-~ 
\frac{2}{m} \delta^{\alpha\beta} \delta^{\sigma\rho} \right) 
\end{equation}  
with the propagator of the field denoted by the same letter. The quantities
$A$, $B$ and $C$ are the momentum independent amplitudes and the exponents are 
defined by 
\begin{equation} 
\alpha ~=~ \mu ~-~ 1 ~+~ \frac{\eta}{2} ~~~,~~~
\beta ~=~ 2 ~-~ \eta ~-~ \chi ~~~,~~~  
\gamma ~=~ 2 ~-~ \eta ~-~ \chi_T  
\end{equation} 
where $d$ $=$ $2\mu$ and $\chi$ and $\chi_T$ are the respective anomalous 
dimensions of the $\sigma$ and $T^{\alpha\beta}$ vertices of (\ref{lag2}). With 
these forms, (\ref{props}) and (\ref{propT}), various critical exponents have 
already been computed in \cite{9,11,17,18,19} in the $1/N$ expansion. Within 
this large $N$ formalism the four fixed points can be identified with whether 
the auxiliary fields propagate or not. For example, at the Gaussian fixed point
the field theory is free and therefore neither $\sigma$ nor $T^{\alpha\beta}$ 
propagate. Likewise the Heisenberg model is clearly recovered from (\ref{lag2})
if $T^{\alpha\beta}$ is set to zero. For the other non-trivial fixed points, 
the unstable antichiral one is characterized by $\sigma$ $=$ $0$ and 
$T^{\alpha\beta}$ $\neq$ $0$ and the stable chiral point is given by $\sigma$ 
$\neq$ $0$ and $T^{\alpha\beta}$ $\neq$ $0$. Therefore, we will compute the 
anomalous dimension of the chiral operator for the full Lagrangian, 
(\ref{lag2}). This bilinear operator, $C^{ab,\alpha\beta}$, is defined by, 
\cite{13}, 
\begin{equation} 
C^{ab,\alpha\beta} ~=~ \phi^{a\alpha} \phi^{b\beta} ~-~ 
\phi^{a\beta} \phi^{b\alpha} 
\end{equation} 
and is traceless. Denoting the anomalous dimension exponent of this operator
by $\eta_c$ then the chiral exponent, $\phi_c$, of \cite{13} is defined by 
\begin{equation} 
\phi_c ~=~ ( 2 ~-~ \eta_c ) \nu 
\label{scal1} 
\end{equation}
where 
\begin{equation} 
\eta_c ~=~ \eta ~+~ \eta_{\cal O} ~. 
\label{scal2} 
\end{equation} 
The exponent $\eta_{\cal O}$ corresponds to the anomalous dimension of the 
bare operator $C^{ab,\alpha\beta}$ itself.  

In the large $N$ determination of the critical exponent of $C^{ab,\alpha\beta}$
the operator is inserted in a two point function where the lines of the graphs
are replaced by the massless propagators (\ref{props}). As we are interested
in $\eta_{\cal O}$ at $O(1/N^2)$ there is only one topology at $O(1/N)$ and
six basic topologies at $O(1/N^2)$. By topology we mean the independent type
of diagrams built from a $\sigma \phi^2$ style vertex. Of these $O(1/N^2)$
topologies four are $2$ loop and two are $3$ loop. This mixture in number of
loops is a consequence of the way one counts powers of $1/N$. In the critical
point method each combination of amplitudes $z$ $=$ $A^2B$ and $y$ $=$ $A^2C$
counts $O(1/N)$ whilst a closed $\phi^{a\alpha}$ field loop is $O(N)$. 
Ordinarily this would mean that there are topologies up to {\em five} loops.
However, due to the traceless nature of $C^{ab,\alpha\beta}$, diagrams where 
the operator is inserted in a closed $\phi^{a\alpha}$ loop yield a zero group
theory factor. In addition to substituting the critical propagators into the
contributing Feynman diagrams a large $N$ regularization needs to be 
introduced. This is provided by the formal shift $\chi$ $\rightarrow$ $\chi$
$+$ $\Delta$, $\chi_T$ $\rightarrow$ $\chi_T$ $+$ $\Delta$. Therefore, each
diagram when computed will involve double and simple poles in the parameter
$\Delta$ which plays a similar role to $\epsilon$ in dimensional 
regularization. These are renormalized in a fashion akin to conventional 
perturbation theory. Consequently the scaling behaviour of the remaining finite
Green's function is exploited to deduce $\eta_{\cal O}$ at $O(1/N^2)$, 
\cite{19}. This method which we have summarized has been used to evaluate a 
variety of operator dimensions in several models. For instance, we have written
a computer algebra programme in the symbolic manipulation language {\sc Form}, 
\cite{21}, to compute $\eta_{\cal O}$. As a non-trivial check on its 
correctness it has been used to re-evaluate the exponent $\chi$ at $O(1/N^2)$ 
at the stable chiral fixed point. Moreover, we have used the same programme to 
compute the crossover exponent at $O(1/N^2)$ at the Heisenberg fixed point 
where the resulting exponent agreed with {\em four} loop $\MSbar$ perturbation 
theory, \cite{20}. 

The result of our computation is 
\begin{eqnarray} 
\eta_{\cal O} &=& \frac{\mu\eta_1}{(m+1)(\mu-2)N} \nonumber \\
&& +~ \left[ \left( \frac{\mu^2(2\mu-1)}{2(m+1)(\mu-2)^2} ~-~ 
\frac{\mu(2\mu^2-5\mu+4)}{2(\mu-2)^2} \right) v^\prime ~-~ \mu ~+~ 
\frac{4}{(\mu-2)} ~+~ \frac{3}{(\mu-2)^2} \right. \nonumber \\ 
&& \left. ~~~~~+~ \frac{2}{(\mu-2)^3} ~-~ \frac{3}{2(\mu-1)} ~+~ 
\frac{\mu^2(2\mu^3-9\mu^2+9\mu-1)}{2(m+1)(\mu-1)(\mu-2)^3} \right] 
\frac{\eta_1}{(m+1)N^2} ~+~ O \left( \frac{1}{N^3} \right) \nonumber \\  
\end{eqnarray} 
where we use the same notation and definitions of \cite{16,17} with  
\begin{equation} 
v^\prime ~=~ \psi(2\mu-2) ~+~ \psi(2-\mu) ~-~ \psi(\mu-2) ~-~ \psi(2) 
\end{equation} 
and $\psi(x)$ $=$ $d\ln\Gamma(x)/dx$ where $\Gamma(x)$ is the Euler gamma
function. The anomalous dimension $\eta$ is given by 
\begin{equation} 
\eta ~=~ \sum_{i=1}^\infty \frac{\eta_i}{N^i} 
\end{equation} 
where, \cite{11}, 
\begin{equation} 
\eta_1 ~=~ -~ \frac{2(m+1)\Gamma(2\mu-2)} 
{\Gamma(\mu+1)\Gamma(\mu-1)\Gamma(\mu-2)\Gamma(2-\mu)} ~.  
\end{equation} 
Using the scaling relations (\ref{scal1}) and (\ref{scal2}), we find 
\begin{eqnarray} 
\phi_c &=& \frac{1}{(\mu-1)} ~+~ \frac{\mu(3m+2)\eta_1}{2(m+1)(\mu-1)(\mu-2)N} 
\nonumber \\
&& +~ \left[ \left( \frac{3\mu^2(5\mu^2-14\mu+10)}{2(\mu-1)(\mu-2)^3} ~+~ 
\frac{3\mu^2(m+3)(3\mu-4)}{2(m+1)^2(\mu-2)^3} \right) R_1 \right. \nonumber \\ 
&& \left. ~~~~~-~ \frac{\mu^2(m^2+3m+4)(2\mu-3)^2}{2(m+1)^2(\mu-1)(\mu-2)^3} 
\left[ R_2 ~+~ R_3^2 \right] \right. \nonumber \\  
&& \left. ~~~~~+~ \left( \mu(6\mu-7)(3\mu-5) ~+~ 
\frac{2(2\mu^3-11\mu^2+12\mu-2)}{(m+1)} \right. \right. \nonumber \\
&& \left. \left. ~~~~~~~~~~~~+~ \frac{\mu(2\mu^2-27\mu+29)}{(m+1)^2} \right) 
\frac{\mu v^\prime}{4(\mu-1)^2(\mu-2)^2} \right. \nonumber \\
&& \left. ~~~~~-~ \frac{1}{2} ~+~ \frac{25}{4(\mu-2)} ~-~ 
\frac{23}{(\mu-2)^2} ~-~ \frac{7}{(\mu-2)^3} ~-~ \frac{8}{(\mu-1)} \right.
\nonumber \\
&& \left. ~~~~~+~ \frac{7}{4(\mu-1)^2} ~+~ \frac{3}{2(\mu-1)^3} ~-~ 
\frac{\mu(27\mu^3-83\mu^2+63\mu-6)}{4(m+1)(\mu-1)^2(\mu-2)^3} \right. 
\nonumber \\
&& \left. ~~~~~-~ \frac{\mu^2(6\mu^3+9\mu^2-67\mu+49)} 
{4(m+1)^2(\mu-1)^2(\mu-2)^3} \right] \frac{\eta_1^2}{N^2} ~+~ O \left( 
\frac{1}{N^3} \right) 
\label{phic} 
\end{eqnarray}  
where 
\begin{eqnarray} 
R_1 &=& \psi^\prime(\mu-1) ~-~ \psi^\prime(1) \nonumber \\  
R_2 &=& \psi^\prime(2\mu-3) ~-~ \psi^\prime(2-\mu) ~-~ \psi^\prime(\mu-1) ~+~
\psi^\prime(1) \nonumber \\ 
R_3 &=& \psi(2\mu-3) ~+~ \psi(2-\mu) ~-~ \psi(\mu-1) ~-~ \psi(1) ~. 
\end{eqnarray} 
For completeness we note that at the stable chiral fixed point, \cite{11}, 
\begin{eqnarray} 
\eta_2 &=& \left[ \left( \frac{1}{2} ~-~ \mu ~-~ \frac{1}{(\mu-2)} ~-~ 
\frac{\mu(m+3)(2\mu-1)}{2(m+1)^2(\mu-2)} \right) v^\prime \right. \nonumber \\ 
&& \left. ~~-~ \frac{(\mu-1)}{2\mu} ~-~ \mu ~-~ \frac{2}{(\mu-2)} ~+~ 
\frac{1}{2(\mu-1)} ~-~ \frac{\mu(m+3)(2\mu^2-5\mu+1)}{2(m+1)^2(\mu-2)^2} 
\right] ~.  
\end{eqnarray}  
The expression for $\phi_c$ at $O(1/N^2)$ represents the main result of this
article and to $O(1/N)$ it is in exact agreement with equation (11) of 
\cite{9}. It can be expanded in $d$ $=$ $4$ $-$ $2\epsilon$ dimensions in 
preparation for comparison with explicit $\MSbar$ perturbative calculations
when they become available. Therefore, we have 
\begin{eqnarray} 
\phi_c &=& 1 ~+~ \epsilon ~+~ \epsilon^2 ~+~ \epsilon^3 ~+~ \epsilon^4 ~-~ 
(3m+2) \left[ \epsilon ~-~ \epsilon^3 ~-~ 2( 1 - \zeta(3) )\epsilon^4  
\right] \frac{1}{N} \nonumber \\ 
&& +~ \left[ ~-~ (3m^2-8m-44)\epsilon ~+~ \frac{1}{2}(24m^2-25m-207)\epsilon^2 
\right. \nonumber \\
&& \left. ~~~~~~-~ \left( 12(2m^2+7m+11)\zeta(3) + \frac{1}{4} (20m^2+123m+339) 
\right) \epsilon^3 \right. \nonumber \\
&& \left. ~~~~~~+~ \left( 80(m^2+3m+4)\zeta(5) - 18(2m^2+7m+11)\zeta(4) 
+ 4(7m^2+58m+149)\zeta(3) \right. \right. \nonumber \\
&& \left. \left. ~~~~~~~~~~~~-~ \frac{1}{8}(218m^2+411m+273) \right) \epsilon^4 
\right] \frac{1}{N^2} ~+~ O\left(\frac{\epsilon^5}{N^3} \right) 
\end{eqnarray}  
for arbitrary $m$ where the order symbol indicates the terms omitted in the
independent expansions in powers of $\epsilon$ and $1/N$ and $\zeta(n)$ is the
Riemann zeta function. For $m$ $=$ $2$ this reduces to 
\begin{eqnarray} 
\phi_c &=& 1 ~+~ \epsilon ~+~ \epsilon^2 ~+~ \epsilon^3 ~+~ \epsilon^4 ~-~ 
8 \left[ \epsilon ~-~ \epsilon^3 ~-~ 2( 1 - \zeta(3) )\epsilon^4 \right] 
\frac{1}{N} \nonumber \\ 
&& +~ \left[ 48\epsilon ~-~ \frac{161}{2}\epsilon^2 ~-~ \left( \frac{665}{4} 
+ 396\zeta(3) \right) \epsilon^3 \right. \nonumber \\
&& \left. ~~~~+~ \left( 1120\zeta(5) - 594\zeta(4) + 1172\zeta(3) 
- \frac{1976}{8} \epsilon^4 \right) \right] \frac{1}{N^2} ~+~ 
O\left(\frac{\epsilon^5}{N^3} \right) ~.  
\end{eqnarray}  
The expansion to $O(1/N)$ illustrates a general property of the general
expression for $\phi_c$ at this order. Comparing (\ref{phic}) with the critical
exponent of the corresponding bilinear operator which relates to crossing 
behaviour at the Heisenberg fixed point, \cite{22,23}, it is in exact agreement
for $m$ $=$ $2$, with the differences in values only appearing at $O(1/N^2)$ in
arbitrary $d$. Thus the $\epsilon$ expansions for both exponents at $O(1/N)$ 
will be the same.

Next we can evaluate our expressions for $m$ $=$ $2$ in three dimensions to 
obtain  
\begin{eqnarray} 
\phi_c &=& 2 ~-~ \frac{32}{\pi^2 N} ~-~ \frac{56[2\pi^2 + 7]}{\pi^4N^2} ~+~ 
O \left( \frac{1}{N^3} \right) \nonumber \\ 
\eta_c &=& -~ \frac{232}{3\pi^4N^2} ~+~ 
O \left( \frac{1}{N^3} \right) \nonumber \\ 
\eta_{\cal O} &=& -~ \frac{4}{\pi^2 N} ~-~ \frac{56}{\pi^4N^2} ~+~ 
O \left( \frac{1}{N^3} \right) 
\label{phic3} 
\end{eqnarray} 
where the intermediate exponents are given by 
\begin{eqnarray} 
\eta &=& \frac{4}{\pi^2 N} ~-~ \frac{64}{3\pi^4N^2} ~+~ 
O \left( \frac{1}{N^3} \right) \nonumber \\ 
\nu &=& 1 ~-~ \frac{16}{\pi^2 N} ~-~ \frac{8[21\pi^2 + 88]}{3\pi^4N^2} ~+~ 
O \left( \frac{1}{N^3} \right) ~. 
\label{etanu} 
\end{eqnarray} 
It is interesting to note that there is no $O(1/N)$ correction to $\eta_c$. 
There are two related chiral exponents defined by the scaling relations in
arbitrary dimensions  
\begin{equation} 
\beta_c ~=~ 2 \mu \nu ~-~ \phi_c ~~~,~~~ \gamma_c ~=~ 2\phi_c ~-~ 2\mu \nu ~. 
\label{betgamdef} 
\end{equation} 
Using (\ref{phic3}), (\ref{etanu}) and (\ref{betgamdef}) we find 
\begin{eqnarray} 
\beta_c &=& 1 ~-~ \frac{16}{\pi^2 N} ~-~ \frac{8[7\pi^2 + 39]}{\pi^4N^2} ~+~ 
O \left( \frac{1}{N^3} \right) \nonumber \\  
\gamma_c &=& 1 ~-~ \frac{16}{\pi^2N} ~-~ \frac{8[7\pi^2+10]}{\pi^4N^2} ~+~ 
O \left( \frac{1}{N^3} \right) 
\end{eqnarray} 
in three dimensions. From the results for each of $\phi_c$, $\beta_c$ and 
$\gamma_c$ it is easy to observe that the coefficient of the $O(1/N^2)$ term
is large and therefore each series is diverging. One way to improve the 
convergence is to apply the Pad\'{e}-Borel resummation technique. For a series
of the form $\sum_{n=0}^\infty a_n x^n$ it can be rewritten in terms of the 
associated Borel function
\begin{equation} 
\sum_{n=0}^\infty a_n x^n ~=~ \frac{1}{x} \int_0^\infty \! dt ~ e^{-t/x} 
\sum_{n=0}^\infty \frac{a_n t^n}{n!} 
\end{equation} 
and then the integrand can be replaced by a Pad\'{e} approximant if a limited
number of coefficients $\{a_n\}$ are known. Hence, for $\phi_c$ we have  
\begin{equation} 
\phi_c ~=~ 2N \int_0^\infty \! dt ~ \frac{e^{-Nt}}{\left[ 1 ~-~ a_1 t ~+~ 
(a_1^2 - \half a_2) t^2 \right]} 
\label{phicpb} 
\end{equation} 
where 
\begin{equation} 
a_1 ~=~ -~ \frac{16}{\pi^2} ~~~,~~~ 
a_2 ~=~ -~ \frac{28[2\pi^2+7]}{\pi^4} ~.  
\end{equation} 
We have evaluated this estimate for $\phi_c$ for various values of $N$ and
recorded them in Table 1 together with the estimates for $\beta_c$ and 
$\gamma_c$ using the same method. 
{\begin{table}[ht] 
\begin{center} 
\begin{tabular}{|c||c|c|c|}
\hline
$N$ & $\phi_c$ & $\beta_c$ & $\gamma_c$ \\ 
\hline 
2 & 0.929 & 0.457 & 0.473 \\ 
3 & 1.131 & 0.558 & 0.574 \\ 
5 & 1.377 & 0.682 & 0.695 \\ 
6 & 1.457 & 0.723 & 0.734 \\
8 & 1.571 & 0.781 & 0.790 \\  
16 & 1.775 & 0.885 & 0.890 \\ 
\hline
\end{tabular} 
\end{center} 
\begin{center} 
{Table 1. Pad\'{e}-Borel estimates of chiral exponents in three dimensions.} 
\end{center} 
\end{table}}  

One has to be careful in interpreting the results in Table 1. Unlike the widely
studied Heisenberg model, it is believed that for low values of $N$ the stable
chiral fixed point of (\ref{lag1}) is not {\em analytically} connected to the 
one used in the large $N$ expansion. Indeed not only does the work of \cite{11}
suggest this but recent exact renormalization group analysis corroborates this 
position, \cite{24,25}. In other words there is evidence that the large $N$ 
expansion is not predictive for $N$ $=$ $2$ and $3$, \cite{24,25}. Nevertheless
if one sets this point of view aside for $N$ $=$ $2$ and $3$ then the results 
of Table 1 follow from (\ref{phicpb}). On the other hand the values for say $N$ 
$=$ $6$ and larger would be expected to be more reliable. Though there are no 
experimental or theoretical results to compare with in these cases. 

To conclude with we have provided the $O(1/N^2)$ expressions for the chiral
exponents of the $O(N)$ $\times$ $O(m)$ $\phi^4$ theory in arbitrary dimensions
at the stable chiral fixed point. At $O(1/N)$ the result for $m$ $=$ $2$ is in
{\em exact} agreement with the expression of the crossover exponent at the
Heisenberg fixed point in the same model. However, in three dimensions for the
specific case $m$~$=$~$2$, the $O(1/N^2)$ corrections are large and therefore 
the series is not converging quickly. Indeed comparison with the crossover 
exponent, \cite{20}, shows that the convergence for the $m$~$=$~$2$ chiral 
exponent is slower and gets slower as $m$ increases.

\end{document}